\magnification=1200
\hsize =6.0 true in
\vsize = 8.0 true in
\hoffset=.375 true in
\voffset=.5true in

\font\mysmall=cmr8 at 8 pt

\def\f{\bf F}

\def\h{\bf H}

\def\z{\bf Z}

\def\p{\bf P}
\def\r{\bf R}
\def\dd{\cal D}

\def\ll{\cal L}

\def\ll{\cal L}
\def\ss{\cal S}

\def\picture #1 by #2 (#3){
		\vbox to #2{
		\hrule width #1 height 0pt depth 0pt
		\vfill
		\special {picture #3}}}
\font\teneufm eufm10 
\font\seveneufm eufm7 
\font\fiveeufm eufm5
\newfam\eufm
\textfont\eufm\teneufm
\scriptfont\eufm\seveneufm
\scriptscriptfont\eufm\fiveeufm

\noindent
\vskip 0.7 true in\noindent
\centerline {\bf   REMARKS ON SOME APPLICATIONS OF SKOROKHOD SPACE}
\bigskip\centerline
{\bf IN QUANTUM MECHANICS} \vskip 0.7 true
in\noindent \centerline {V. S. VARADARAJAN}
\bigskip
\centerline{Department of Mathematics, University of California}
\centerline {Los Angeles, CA 90095--1555, USA} 
\centerline {\mysmall e-mail :  vsv@math.ucla.edu}
\vskip 0.7 true in
\hfill {\it Dedicated to A. V. Skorokhod }
\vskip 0.7 true in\noindent
{\bf Abstract.\/} {\mysmall This paper discusses the role of the Skorokhod space
and the convergence of probability measures on it in some recent studies of the foundations of
quantum mechanics, both in the conventional setting over the real number field and in the more
speculative one of nonarchimedean local fields.} \vskip 0.5 true in\noindent  {\bf AMS subject
classification (1991).\/}  81 S 40, 60 F 99.  \vskip 0.9 true in\noindent  {\bf 1. Introduction.} 
\medskip\noindent It is a pleasure and honour for me to have been asked to contribute to to the
collection of articles marking the 40$^{\rm th}$ anniversary of the discovery of the {\it Skorokhod
topology\/} by Professor A. V. Skorokhod. I was a graduate student in Probability theory at the
Indian Statistical Institute, Calcutta, India in 1956, and still remember vividly the surprise and
excitement of myself and of my fellow students when the first papers on the subject by Skorokhod
himself$^1$ and Kolmogorov$^2$ appeared. It was clear from the beginning that the space $D$
with its Skorokhod topology would play a fundamental role in all problems where limit theorems
involving stochastic processes whose paths are not continuous (but are allowed to have only
discontinuities of the first kind) were involved.  \medskip The present paper is a brief review of the
use of the Skorokhod space and convergence of probability measures on it in some recent  studies
of quantum systems over fields and rings not only over the reals, but also over $p$-adic
fields$^{3,4}$. The first application I discuss is to the approximation of usual quantum systems by
finite quantum systems$^3$. The second$^4$ is a discussion of a path integral formalism
applicable to a class of $p$-adic Schr\"odinger equations; the corresponding probabi;ity measure 
comes from a stochastic process with independent increments and  
is defined on the Skorokhod space of functions on $[0, \infty )$ with values in a finite dimensional
vector space over a nonarchimedean local field. This stochastic process and the associated
measure therefore play the same role in the study of these $p$-adic Schr\"odinger equations as
the brownian motion in the theory of the usual Schr\"odinger equations. 
 \bigskip \noindent
{\bf 2. Finite approximations of usual quantum systems.\/} The idea of studying finite quantum
systems and their limiting forms goes back to Weyl$^5$ in the 1930's and Schwinger$^6$ in the
1960's, and has still retained great interest$^7$. For both Schwinger and Weyl one of the themes
was to approximate quantum systems over ${\r}$ by finite quantum systems obtained by replacing
${\r}$ with the cyclic group ${\z}_N={\z}/N{\z}$ for $N$ large (this is also the basic idea in the
so-called theory of the fast Fourier transform), identifying ${\z}_N$ with the grid  $\{0, \pm
\varepsilon, \pm 2\varepsilon, \dots ,\pm k\varepsilon\}$ where $N=2k+1$ and $\varepsilon =(2\pi
/N)^{1/2}$. Weyl was interested only in the kinematics while Schwinger was interested in the
dynamics also. Schwinger introduced the position coordinate $q_N$ as the multiplication by the
function $k\varepsilon \mapsto k\varepsilon $ on the grid, and the momentum coordinate $p_N$ as
the {\it Fourier transform\/} of $q_N$ on the finite group ${\z}_N$ using the identification above.
Schwinger's principle was that the finte dimensional operator $H_N^{(s)}=(1/2)p_N^2+V(q_N)$ is a
very good approximation to the energy operator $H=(1/2)p^2+V(q)$ for large $N$. Numerical work
for the case of the harmonic oscillator showed that this was true$^3$, and the question naturally
arose if this could be substantiated by a limit theorem. In$^3$ it was shown that, in arbitrary
dimension $d$ and for potentials $V$ which go to infinity faster than $\log r$ at infinity on ${\r}^d$,
we have $$
||e^{-tH_N^{(s)}}-e^{-tH}||_1\longrightarrow 0 \qquad (N\to \infty )
$$
where $||{\cdot}||_1$ is the trace norm (the condition on $V$ insures that the operators
$e^{-tH}$ are of trace class for every $t>0$).

The method of proving this theorem is to use the Feynman-Kac formula$^8$ for the propagators
of the Hamiltonian $H$. Such a formula is not available for the approximating 
Schwinger Hamiltonian $H_N^{(s)}$; but, if one replaces the free Hamiltonian by a second
difference operator which is the discrete analogue of the Laplacian,  then one has
such a formula. One can call such Hamiltonians {\it stochastic\/} because the measure on the path
space comes from a stochastic process with independent increments, namely the random walk, on
the lattice ${\ll}_N=(\varepsilon {\z}^d)$. In the case of the finite approximation when the infinite
lattice is truncated to a finite one, the path space measure still exists, but is now assocaited to a
random walk with some boundary conditions that keep the walk inside the finite grid ${\ll}_N^\ast $.
It is not difficult to show that the Schwinger Hamiltonian is a better approximatioin than the
stochastic Hamiltonian and so it is enough to establish the limit theorem for the stochastic ones.
We shall denote these by $H_N$ in the case of the infinite lattice and $H_N^\ast $ in the case of
the finite lattice. 

For the continuum limit the path integral defining the propagator is with respect to the
so-called brownian bridges, namely the measures ${\p}^t_{x,y}$ defined by the
conditional brownian motion starting from $x$ at time $0$ and exiting at time $t$ through $y$.
But, for the approximating processes, the measures are defined only on step functions with values
in the lattices ${\ll}_N, {\ll}_N^\ast $. It is therefore essential, since one wants to discuss the
approximation at the level of the probabiltiy measures on path spaces, to have all the measures
defined on a single space. This has to be the Skorokhod space ${\dd}_t$ of functions on $[0, t]$
with values in ${\r}^d$ with discontinuities only of the first kind. 

The fundamental result that allows one to prove the approximation theorem is the following local
limit theorem on the Skorokhod space. Let ${\p}^t_{N,a,b}$ be the conditional probabilty measure
on ${\dd}_t$ for the random walk on the approximating lattice ${\ll}_N$ that starts from $a\in
{\ll}_N$ at time $0$ and exits from $b\in {\ll}_N$ at time $t$. Then 
\bigskip\noindent
{\bf Theorem\/} {\it Fix $x, y\in {\r}^d$ and let $a, b\in {\ll}_N$ vary in such a manner that $a\to x,
b\to y$ as $N\to \infty $. Then  $$
 {\p}^t_{N,a,b}\Longrightarrow {\p}^t_{x,y}
$$
in the sense of weak convergence of measures on ${\dd}_t$.}
\bigskip
Let us now recall that the operators $e^{-tH}$ and $e^{-tH_N}$ are integral operators with
kernels $K_t, K_{N,t}$ where
$$
\eqalign {K_t(x, y)&=\int _{{\dd}_t} e^{-\int _0^t V(\omega (s)) ds } d{\p}^t_{x,y}(\omega )\cr 
 K_{N,t}(x, y)&=\int _{{\dd}_t} e^{-\int _0^t V(\omega (s)) ds } d{\p}^t_{N,x,y}(\omega )\cr }
$$
The traces of these integral operators are calculated by integrating the kernels on the
diagonal. The theorem above now leads to the limit formula
$$
Tr (e^{-tH_N})=\sum _{a\in {\ll}_N} K_{N, t}(a,a) \to Tr (e^{-tH})=\int _{{\r}^d} K_t(x, x) dx
$$
The second step is  then to show that the trace limit relation continues to hold on going from
${\ll}_N$ to ${\ll}_N^\ast $. This can be done, and one has the following limit formula:
$$
Tr (e^{-tH_N^\ast })=\sum _{a\in {\ll}_N^\ast  } K_{N, t}(a,a) \to Tr (e^{-tH})=\int _{{\r}^d} K_t(x, x) dx
$$
The required approximation of $e^{-tH}$ by $e^{-tH_N^\ast }$ in trace norm then follows from some
standard arguments from functional analysis.

The limit theorem and its consequence require extensive use of techniques that are basic  to
the theory of the Skorokhod spaces and are discussed in detail in$^3$.          \bigskip\noindent 
{\bf 3. $\bf p$-adic Schr\"odinger equations and path integral representations for their propagators
in imaginary time.\/}  \medskip\noindent  Already in the 1970's and in fact much earlier even
there was interest in understanding the structure of quantum mechanical theories over
nonarchimedean local fields and even discrete structures like finite fields$^9$. In recent years this
interest has deepened, and mathematical and physical questions which may be viewed as the
nonarchimedean counterparts of well-known quantum mechanical questions have begun to be
studied over nonarchimedean fields$^{10}$. In this section I shall discuss briefly one such aspect of
$p$-adic analysis, namely, Schr\"odinger equations over $p$-adic fields; the proofs of the
statements made here will appear elsewhere$^4$.

Let $K$ be any nonarchimedean local
field of arbitrary characteristic and $D$ a division algebra of finite dimension over $K$. We shall 
assume  that $K$ is the center of $D$; this is no loss of generality since we may always replace
$K$ by the center of $D$. Let $dx$ be a Haar measure on $D$ and $|\cdot|$ the usual  modulus
function on $D$: $$ d(ax)=|a| dx \quad (a\not=0),\qquad |0|=0 $$ It is then immediate that $|\cdot |$
is a multiplicative norm which is  ultrametric (i.e., $|x+y|\le \max (|x|, |y|)$) that induces the original
topology.

Let $F$ be a left vecor space of finite dimension over $D$. By a {\it $D$-norm\/} on $F$ is meant a
function $|\cdot|$ from $F$ to the nonnegative reals such that  \medskip\itemitem {(i)} $|v|=0$ if and
only if $v=0$  \itemitem {(ii)} $|av|=|a||v|$ for $a\in D$ and $v\in F$  \itemitem {(iii)} $|\cdot|$
satisfies the ultrametric inequality, i.e.,  $$ |u+v|\le \max (|u|, |v|)\quad (u, v\in F) $$
\medskip\noindent The norm on the dual $F^\ast $ of $F$  is a $D$-norm. If we identify $F$ with
$D^n$ by choosing a basis,  and define, for suitable constants $a_i>0$, $$ |v|=\max _{1\le i\le
n}(a_i|v_i|)\quad (v=(v_1, v_2, \dots , v_n)) $$
it is immediate that $|\cdot|$ is a $D$-norm. It is known that
every $D$-norm is of this form. In particular all these norms induce the same locally compact
topology on $F$.

For $x\in F, \xi \in F^\ast $, let us write $x\xi $ for the value of $\xi $ at $x$. If $\chi $ is a nontrivial
additive character on $D$, then  $\psi _\xi (x\longmapsto \chi (x\xi ))$ is an additive character of
$F$, every additive character of $F$ is of this form, and the map $\xi \longmapsto \psi _\xi $ is an
isomorphism of topological groups from $F^\ast $ to $\hat F$, the dual group of $F$. By ${\ss}(F)$
we denote the Schwartz-Bruhat space of complex-valued locally constant functions with compact
supports on $F$. Let $dx$ be a Haar measure on $F$. Then ${\ss}(F)$ is dense in $L^2(F, dx)$,
and the Fourier transform ${\f}$ is an isomorphism of ${\ss}(F)$ with ${\ss}(F^\ast )$, defined by
$$
{\f}(g)(\xi )=\int \chi (x\xi )g(x)dx\quad (\xi \in F^\ast )
$$ For a unique choice of Haar measure on $F^\ast $ we have, for all $g\in
{\ss}(F)$,
$$
g(x)=\int \chi (-x\xi ){\f}g(\xi )d\xi \quad (x\in F)
$$
The measures $dx$ and $d\xi $ are then said to be {\it dual\/} to each other. For all of this,
see$^{11}$.

It is natural to call $p$-adic Schr\"odinger theory the study of the
spectra and semigroups generated by operators in $L^2(F)$  where $F$ is a finite dimensional
vector space over $D$, of the form $$ H=H_0 + V $$ Here $H_0$ is a pseudodifferential operator
and $V$ is a multiplication operator. The simplest examples of $H_0$ are as follows. We write
$M_b$ for multiplication by $|x|^b (b>0)$ in  $$ {\h}=L^2(F) $$   
and, denoting by ${\f}$ the Fourier transform on ${\h}$, we put
$$
{\Delta }_{F,b}={\f}M_b{\f}^{-1}
$$
The Hamiltonian will then be of the form
$$
H_{F,b}={\Delta }_{F,b}+V
$$
It is clear that over the field of real numbers and for $b=2$ the operator ${\Delta }_{F,b}$ is 
just $-\Delta $ where $\Delta $ is the Laplace operator. The Hamiltonians $H_{F,b}$ are thus the 
counterparts over $D$ to the usual
ones that appear in the conventional Schr\"odinger equations.

I shall now indicate how a path integral representation can be given for the propagators in
iamginary time for the Hamiltonians $H$ defined above in the nonarchimedean context. The key
is the following.
\bigskip\noindent
{\bf Proposition\/} {\it Fix $t>0$ and $b>0$ and let $F$ be a $n$-dimensional left vector space
over $D$ with a $D$-norm $|\cdot|$. Then the function $\varphi $ on $V^\ast $ defined by  $$
\varphi (\xi )= \exp (-t|\xi |^b) \quad (\xi \in F^\ast ) $$   is in $L^m(F, d\xi )$ for all $m\ge 1$
and is positive definite.  If we denote by $f_{t,b}$ the (continuous) probability density on $F$ whose
Fourier transform is $\varphi $, then $f_{t,b}$ is $>0$ everywhere. Moreover (i) $0<f_{t,b}(x)\le
f(0)\le A\ t^{-n/b}$ for all $t>0$, $A$ being a constant $>0$ not depending on $t$ (ii) For $0\le k<b$
we have, for all $t>0$ and a constant $A>0$ independent of $t$,  $$
 \int _F|x|^k f_{t,b}(x) dx \le A\ t^{k/b}$$}
\bigskip
It follows from this that the $(f_{t,b})_{t>0}$ form
a continuous convolution semigroup of probability measures which goes to the Dirac delta measure
at $0$ when $t\to 0$. Hence for any $x\in F$ one can associate a separable $F$-valued
stochastic process with independent increments $(X(t))_{t\ge 0}$ with $X(0)=x$, such that
$f_{t,b}$ is the density of the distribution of $X(t+u)-X(u)$ for any $t>0, u\ge 0$. 

Let $D([0, \infty ) : M)$ be
the space of right continuous functions on $[0, \infty )$ with values in the complete separable metric
space $M$ having only discontinuities of the first kind. For any $T>0$ we write  $D([0,T] : M)$ for
the analogous space of right continuous functions on $[0,T)$ with values in the complete separable
metric space $M$ having only discontinuities of the first kind, and left continuous at $T$. Then 
one can prove that the $X$-process has sample paths in the Skorokhod space $D([0, \infty ) :
F)$. More precisely we have .  \bigskip\noindent
{\bf Theorem \/} {\it There are unique families of probability measures ${\p}^b_x$ on 
$D([0, \infty ) : F)$ and ${\p}^{T,b}_{x,y}
(x, y\in F)$  on $D([0, T] : F)$ , continuous with respect to $(x,y)$, such that 
${\p}^b_x$ is the measure of the $X$-process that starts from $x$ at time $t=0$, and
${\p}^{T,b}_{x,y}$ is the probability  measure for the $X$-process that starts from $x$ at time $t=0$
and is conditioned to pass through $y$ at time $t=T$.} \bigskip\noindent
{\bf  Feynman--Kac propagator for $e^{-tH_{F,b}} (t>0)$\/} From now on one can use standard
arguments$^{8}$, when $V$ is bounded below and $H_{F,b}$ is essentially self-adjoint on
${\ss}(F)$, to show that the operator $e^{-tH_{F,b}} (t>0)$  is an integral operator in $L^2(F)$ with
kernel $$ K_{t,b}(x : y) \qquad (x,y\in F)
$$ which is represented by the following integral on the space ${\dd}_t=D([0, t] : F)$ :
$$
K_{t,b}(x : y)=\int _{{\dd}_t} \exp \left ( -\int _0^t V(\omega (s))ds\right )dP^{t,b}_{x,y}(\omega )
{\cdot} f_{t,b}(x-y)$$
\vskip 0.7 true in  \centerline {\bf
References\/} \bigskip
 \item {1.} Skorokhod, A. V., {\it Dokl. Akad. Nauk. SSSR\/}, {\bf 104} (1955), 364; {\bf 106},
(1956), 781.   . \smallskip  
\item {2.} Kolmogorov, A. N., {\it Theor. Prob. Appl.\/} {\bf 1},(1956), 215. 
 \smallskip   
\item {3.} Digernes, T., Varadarajan, V. S., and Varadhan, S. R. S., {\it Rev. Math. Phys.\/} {\bf 6}
(1994), 621. 
\smallskip
\item {4.} Varadarajan, V. S., {\it A path integral formalism for a class of $p$-adic Schr\"odinger
equations. (In preparation)\/} 
\smallskip
\item {5.} Weyl, H., {\it Theory of Groups and Quantum Mechanics\/},
Dover, 1931, Ch. III, \S16, Ch. IV, \S\S 14, 15.
 \smallskip
\item {6.} Schwinger, J., {\it Quantum Kinematics and Dynamics\/}, W. A.
Benjamin, 1970. \smallskip   
\item {7.} Varadarajan, V. S., {\it Lett. Math. Phys.\/} {\bf 34}
(1995), 319. \smallskip
\item {} Husstad, E., {\it Thesis\/}, University of Trondheim, 1991/92.
\smallskip
\item {} Stovicek, P., and Tolar, J., {\it Quantum mechanics in
a discrete space-time\/} Rep. Math. Phys. {\bf 20} (1984), 157. 
\smallskip 
\item {8.} Simon, B., {\it Functional integration and quantum physics\/}, Academic Press,
1979.  \smallskip
\item {9.} Ulam, S. {\it Sets,
Numbers, and Universes, Selected Works\/}, MIT Press 1974. See paper [86] (p 265) with
commentary by E. Beltrametti, p 687.  
\smallskip \item {} Beltrametti, E. G. {\it Can a finite geometry
describe the physical space-time?\/}, Atti del convegno di geometria combinatoria e sue
applicazioni, Perugia 1971.\smallskip \smallskip  
\item {10.} Vladimirov, V. S., and Volovich, I., {\it Lett.
Math. Phys.\/} {\bf 18} (1989), 43 \smallskip
\item {} Vladimirov, V. S., {\it Leningrad Math. J\/} {\bf 2} (1991), 1261.  \smallskip 
\item {} Parisi, G., {\it Modern Phys. Lett. A3\/} (1988), 639
\smallskip
\item {} Meurice, Y., {\it Phys. Lett. B\/} {\bf
104} (1990), 245.\smallskip
\item {}  Zelenov, E. I., {\it J. Math. Phys.\/} {\bf 32} (1991), 147.    \smallskip 
\item {} Brekke, L., and Freund, P. G. O., {\it Physics Reports\/} {\bf 233} (1993), 1. 
\smallskip 
\item {11.} Weil, A., {\it Basic Number Theory\/}, Springer, 1961
\smallskip

\bye